\documentclass[10pt]{iopart}
\usepackage{graphicx}
\usepackage{iopams}

\newcommand{\half}{{\textstyle \frac{1}{2}}}
\newcommand{\sumprime}{\! \! \!\!\!\!\!  '  \; \;\;\;}%

\begin{document}
\jl{1} 

\title[Sum rules for ionic mixtures]
{Sum rules for correlation functions of ionic mixtures in arbitrary
dimension $d\geq 2$}

\author{L G Suttorp}

\address{Instituut voor Theoretische Fysica, Universiteit van Amsterdam,
Valckenierstraat 65, 1018 XE Amsterdam, The Netherlands}

\begin{abstract}
The correlations in classical multi-component ionic mixtures with spatial
dimension $d\geq 2$ are studied by using a restricted grand-canonical
ensemble and the associated hierarchy equations for the correlation
functions. Sum rules for the first few moments of the two-particle
correlation function are derived and their dependence on $d$ is
established. By varying $d$ continuously near $d=2$ it is shown how the sum
rules for the two-dimensional mixture are related to those for mixtures at
higher $d$.
\end{abstract}   


\maketitle

\section{Introduction}\label{sec1}
The statistical equilibrium properties of classical many-particle systems
 with long-range forces have been the subject of an extensive literature
 (for reviews see \cite{A86}-\cite{BM99}). The simplest models with
 long-range interactions are Coulomb systems consisting of point particles
 with charges of the same sign that move in an inert uniform background of
 opposite sign. For these systems no collapse of particles can occur and
 stability is guaranteed at all densities and temperatures. Both the
 one-component plasma, also known as jellium, and ionic mixtures of
 particles with different charges and masses fall in this class.  An
 important tool in the analysis of the equilibrium behaviour of these
 systems is furnished by the set of correlation functions and the
 associated Ursell functions. The first few moments of the latter satisfy
 sum rules, which are essential for the description of the large-scale
 fluctuations of local densities.

In studying one-component plasmas and ionic mixtures it has been found that
the dimension $d$ of space in which these systems are embedded plays a
remarkable role. It turns out that several properties of systems with $d=2$
and $d=3$ (which have mainly been considered) are quite similar, whereas
occasionally the derivation of these properties proceeds along rather
different lines. An example is a recent proof of a second-moment sum rule
for correlations near a guest charge in a two-dimensional one-component
plasma \cite{S07}. Here the use of symmetry properties of the Ursell
functions leads to a short proof \cite{JS08}, whereas in deriving the
analogous sum rule for the three-dimensional case a detailed analysis of
the statistical ensemble properties has to be carried out
\cite{SvW87}. Sometimes the analogy between the two- and three-dimensional
cases gets lost altogether, as seems to be the case for a higher-order sum
rule of the two-dimensional one-component plasma \cite{KMST00}. For
this sixth-moment rule no counterpart at $d=3$ has been found as yet.

The purpose of the present paper is to postpone any choice of dimension and
to derive sum rules that are valid for ionic mixtures in all dimensions
$d\geq 2$. We shall refrain from a discussion of the case $d=1$, as 
periodic oscillations in the density lead to complications in that case
\cite{M88}. We shall concentrate on sum rules for two-particle Ursell
functions. Our unified treatment enables one to clearly see how the
simplifications in the derivation of these sum rules for $d=2$ come about,
and why the proof for $d>2$ (and hence for $d=3$ in particular) is
necessarily more complicated. In the course of our analysis we shall obtain
several new results for a general Coulomb-type system with $d>3$, which has
hardly been discussed in the past \cite{GLM80}-\cite{A88}.
In deriving our results we shall treat $d$ as a continuous variable, as is
standard practice in the theory of phase transitions \cite{WF72} and in
dimensional regularization of quantum field theory \cite{tHV72}. This
method has been used in the context of systems with long-range forces as
well \cite{D76}.

When describing multi-component ionic mixtures attention has to be paid to
a suitable choice of the equilibrium ensemble. As in a previous treatment
\cite{SvW87}, we shall use a restricted grand-canonical ensemble, in
which the fluctuating particle numbers are constrained by stipulating that
the ensuing total charge matches the fixed charge of the inert background.

\section{Ionic mixtures in dimension $d\geq 2$}
We consider a $d$-dimensional multi-component ionic mixture of $s$
components, with label $\sigma=1,\ldots,s$, in a large volume $V$. The
$N_\sigma$ particles of species $\sigma$ carry mass $m_\sigma$ and positive
charge $e_\sigma$. The system is neutral owing to a uniform background with
charge density $- q_v \equiv -\sum_{\sigma}e_{\sigma}N_{\sigma}/V$.

For arbitrary $d$ the potential $\phi$ depending on the distance $r=|{\bf
r}|$ is proportional to $1/r^{d-2}$. It is the solution of the
$d$-dimensional inhomogeneous Laplace equation $\Delta \phi(r)=-\delta({\bf
r})$, with $\Delta$ the $d$-dimensional Laplace operator and $\delta({\bf
r})$ the Dirac delta function in $d$ dimensions. Here it should be noted
that in a space with dimension $d$ the Laplace operator acting on an
isotropic function is given by $r^{-d+1}\; (\partial/\partial r)\;
r^{d-1}\; (\partial/\partial r)$. The explicit form of $\phi(r)$ is
\begin{equation}
\phi(r)=\frac{\Gamma(d/2-1)}{4 \pi^{d/2}}\, \frac{1}{r^{d-2}} + c_d
\label{2.1}
\end{equation}
with $\Gamma(z)$ the gamma function and with $c_d$ an arbitrary additive
constant. We used the fact that the surface of a unit sphere in $d$
dimensions equals $2\; \pi^{d/2}/\Gamma(d/2)$. For $d=3$ the potential has
the form $\phi(r)=1/(4\pi r)$ (at least for $c_d=0$), which corresponds to
the choice of so-called rationalized Lorentz-Heaviside units in
electrodynamics. For $d=2$ the potential $\phi$ that solves the
two-dimensional Laplace equation is logarithmic:
\begin{equation}
\phi(r)=-\frac{1}{2\pi}\; \log(r) +c
\label{2.2}
\end{equation}
with a constant $c$ that can be used to render the argument of the
logarithm dimensionless by writing $c=\log(L)/(2\pi)$ with an arbitrary
length $L$. This potential can be obtained from (\ref{2.1}) by taking the
limit $d\rightarrow 2$, if $c_d$ is chosen as :
\begin{equation}
c_d= -\frac{\Gamma(d/2 -1)}{4 \pi^{d/2}}+c
\label{2.3}
\end{equation}
Indeed, in the limit $d\rightarrow 2$ one finds:
\begin{equation}
\lim_{d\rightarrow 2} \phi(r)=
\lim_{d\rightarrow 2} \frac{\Gamma(d/2-1)}{4 \pi^{d/2}}\, 
\left(\frac{1}{r^{d-2}}-1\right)+c=-\frac{1}{2\pi}\; \log(r) +c
\label{2.4}
\end{equation}
It should be noted that the shift in energy $c_d$ as given by (\ref{2.3})
becomes infinite, when $d$ tends to 2. This does not come as a surprise
since the potential (\ref{2.2}) grows without bound for large $r$, whereas
the potential (\ref{2.1}) for $d>2$ tends to $c_d$ at large $r$. If
desired, one may choose $c_d$ to be given by (\ref{2.3}) for all
$d$. However, we shall see that for $d>2$ many formulas simplify by
choosing $c_d=0$, so that the choice (\ref{2.3}) is somewhat artificial in
that case. For that reason we shall postpone a specific choice of $c_d$ and
leave it arbitrary as yet.

The Hamiltonian of the ionic mixture is the sum of the kinetic energy $T$ and
the potential energy $U$:
\begin{eqnarray}
\fl H=T+U=\sum_{\sigma\alpha} \frac{p_{\sigma\alpha}^2}{2m_\sigma} +
\half \sum_{\sigma_1 \alpha_1,\sigma_2 \alpha_2}\sumprime 
e_{\sigma_1}\, e_{\sigma_2}\; 
\phi(|{\bf r}_{\sigma_1\alpha_1}-{\bf r}_{\sigma_2\alpha_2}|)\nonumber\\
-q_v\sum_{\sigma\alpha} e_{\sigma}\int^V d{\bf r}\; 
\phi(|{\bf r}_{\sigma\alpha} -{\bf r}|)
+\half\, q_v^2\int^V d{\bf r}\; d{\bf r}' \; \phi(|{\bf r} -{\bf r}'|)
\label{2.5}
\end{eqnarray}
The particle $\alpha$ of species $\sigma$ has position ${\bf
r}_{\sigma\alpha}$ and momentum ${\bf p}_{\sigma\alpha}$. The prime at the
summation sign indicates the condition $\sigma_1\alpha_1 \neq
\sigma_2\alpha_2$, so that self interactions among the point particles are
excluded. As said above, the constant $c_d$ in the potential is left
arbitrary for the time being. The integrals representing the interactions
involving the background are taken over the $d$-dimensional volume $V$.

As was shown by Lieb and Narnhofer \cite{LN75} for the one-component plasma
in dimension $d=3$, the potential energy $U$ in $H$ is bounded from below,
so that the stability of the system is warranted in that case. Generalizing
their argument so as to be applicable to a mixture in arbitrary dimension
one may prove stability for any $d\geq 2$, as is shown in appendix A. 

\section{Electrostatic sum rules}
\setcounter{equation}{0} The $k$-particle equilibrium correlation functions
$g^{(k)}_{\sigma_1 \ldots \sigma_k}$ satisfy the BGY hierarchy equations
\cite{M88}
\begin{eqnarray}
\fl \frac{\partial}{\partial {\bf r}_1}\; 
g^{(k)}_{\sigma_1 \ldots \sigma_k}({\bf r}_1, \ldots,{\bf r}_k)=
-\beta\;e_{\sigma_1}\;g^{(k)}_{\sigma_1 \ldots \sigma_k}({\bf r}_1, \ldots,{\bf r}_k)\;
\sum_{j=2}^k e_{\sigma_j}\; \frac{\partial}{\partial {\bf r}_1} 
\phi(|{\bf r}_1-{\bf r}_j|)\nonumber\\
\fl  -\beta\;e_{\sigma_1}\sum_{\sigma_{k+1}} n_{\sigma_{k+1}}
\; e_{\sigma_{k+1}}\int^V d{\bf r}_{k+1}
\left[ g^{(k+1)}_{\sigma_1\ldots\sigma_{k+1}}({\bf r}_1,\ldots,{\bf r}_{k+1})
-g^{(k)}_{\sigma_1\ldots\sigma_{k}}({\bf r}_1,\ldots,{\bf r}_{k})\right]\,
\nonumber\\
\times \frac{\partial}{\partial {\bf r}_1}
\phi(|{\bf r}_1-{\bf r}_{k+1}|)
\label{3.1}
\end{eqnarray}
with $\beta =(k_B T)^{-1}$ the inverse temperature and $n_\sigma=\langle
N_\sigma\rangle/V$ the average particle density of species $\sigma$. The
correlation functions can be expanded in terms of Ursell functions
$h^{(k)}_{\sigma_1 \ldots \sigma_k}$ \cite{dB49,V85}. In particular, the
two-particle Ursell function $h^{(2)}_{\sigma_1\sigma_2}$ is defined as
$g^{(2)}_{\sigma_1\sigma_2}-1$. For large $V$ the Ursell functions are
translationally invariant, so that they depend on the difference between
the positions only. In the following we shall assume that the Ursell
functions satisfy the standard exponential clustering hypothesis, which
implies that they tend to zero faster than any power if the separation
between two positions goes to infinity.

For $k=2$ the hierarchy equation reads in terms of the Ursell functions:
\begin{eqnarray}
\fl \beta e_{\sigma_1}\sum_{\sigma_3} n_{\sigma_3}e_{\sigma_3}\int d{\bf r}_3\; 
h^{(3)}_{\sigma_1\sigma_2\sigma_3}({\bf r}_1,{\bf r}_2,{\bf r}_3)\;
\frac{\partial \phi(r_{13})}{\partial{\bf r}_1}=
-\frac{\partial h^{(2)}_{\sigma_1\sigma_2}({\bf r}_1,{\bf r}_2)}
{\partial{\bf r}_1}\nonumber\\
 -\beta\, e_{\sigma_1}\frac{\partial}{\partial{\bf r}_1}
\sum_{\sigma_3} n_{\sigma_3}e_{\sigma_3}\int d{\bf r}_3\;
h^{(2)}_{\sigma_2\sigma_3}({\bf r}_2,{\bf r}_3)\; 
\phi(r_{13})\nonumber\\
-\beta\, e_{\sigma_1}\, e_{\sigma_2}\; 
h^{(2)}_{\sigma_1\sigma_2}({\bf r}_1,{\bf r}_2)\; 
\frac{\partial\phi(r_{12})}{\partial{\bf r}_1}
-\beta\, e_{\sigma_1}\, e_{\sigma_2}\; \frac{\partial
  \phi(r_{12})}{\partial{\bf r}_1}
\label{3.2}
\end{eqnarray}
with ${\bf r}_{ij}={\bf r}_i-{\bf r}_j$.

The second term at the right-hand side can be rewritten by expanding the
potential in terms of Gegenbauer polynomials.  For $r> r'$
one has \cite{E53}:
\begin{equation}
\phi(|{\bf r}-{\bf r}'|)=\phi(r)+ \frac{\Gamma(d/2 -1)}{4\pi^{d/2}\; r^{d-2}}
\sum_{\ell=1}^{\infty} 
C_{\ell}^{(d-2)/2}(\cos\theta)\; \left(\frac{r'}{r}\right)^\ell
\label{3.3}
\end{equation}
with $d>2$. Here $\theta$ is the angle between ${\bf r}$ and ${\bf
r}'$. For $r< r'$ a similar expansion holds, with ${\bf r}$ and ${\bf r}'$
interchanged. By expanding the potential in this way and using the
orthogonality relation of the Gegenbauer polynomials one may establish the
identity
\begin{equation}
\fl \frac{\partial}{\partial{\bf r}_1} \int d{\bf r}_3\;
h^{(2)}_{\sigma_2\sigma_3}({\bf r}_2,{\bf r}_3)\;
\phi(r_{13})=\frac{\partial\phi(r_{12})}{\partial{\bf r}_1}
\int_{r_{23}<r_{12}} d{\bf r}_3\; 
h^{(2)}_{\sigma_2\sigma_3}({\bf r}_2,{\bf r}_3)
\label{3.4}
\end{equation}
Employing this equality in (\ref{3.2}) and making use of the exponential
clustering properties of the Ursell functions one proves the
perfect-screening condition \cite{GLM80,MY80,BGLeM82,vBF79,GLeM81} for the
two-particle Ursell function of a general $d$-dimensional ionic mixture:
\begin{equation}
\sum_{\sigma_2} n_{\sigma_2}\, e_{\sigma_2} \int d{\bf r}_{2}\; 
h_{\sigma_1\sigma_2}^{(2)}({\bf r}_1,{\bf r}_2) = -e_{\sigma_1}
\label{3.5}
\end{equation}

Similarly, by using the Gegenbauer expansion and the exponential clustering
property one derives from the hierarchy equations for $k=3$ the perfect
screening rules:
\begin{eqnarray}
\fl \sum_{\sigma_3} n_{\sigma_3}\, e_{\sigma_3} \int d{\bf r}_{3}\; 
h_{\sigma_1\sigma_2\sigma_3}^{(3)}({\bf r}_1,{\bf r}_2,{\bf r}_3)
=-(e_{\sigma_1}+e_{\sigma_2})\; h_{\sigma_1\sigma_2}^{(2)}({\bf r}_1,{\bf r}_2)
\label{3.6}\\
\fl \sum_{\sigma_3} n_{\sigma_3}\, e_{\sigma_3} \int d{\bf r}_{3}\; 
h_{\sigma_1\sigma_2\sigma_3}^{(3)}({\bf r}_1,{\bf r}_2,{\bf r}_3)\;  r_{13}^\ell\;
C_{\ell}^{(d-2)/2}(\cos\theta)=\nonumber\\
=-\frac{(d-2)_\ell}{\ell!}\; e_{\sigma_2}\;
r_{12}^\ell \; h_{\sigma_1\sigma_2}^{(2)}({\bf r}_1,{\bf r}_2) \quad (\ell=1,2,\ldots)
\label{3.7}
\end{eqnarray}
with $\theta$ the angle between ${\bf r}_{12}$ and ${\bf r}_{13}$, and with
$(x)_n=x(x+1)\cdots(x+n-1)$ the Pochhammer symbol.

For $d=3$ the Gegenbauer polynomials in (\ref{3.7}) reduce to Legendre
polynomials, so that we recover one of the well-known perfect-screening
rules for a three-dimensional ionic mixture \cite{SvW87,BGLeM82,FM84}. To
derive the analogous identity for $d=2$ we use for $\ell\geq 1$
\cite{E53}:
\begin{equation}
\lim_{\lambda\rightarrow 0} \frac{1}{\lambda}\; C^{\lambda}_{\ell}(x)= \frac{2}{\ell}\;
T_{\ell}(x)
\label{3.8}
\end{equation}
with $T_\ell(x)$ the Chebyshev polynomials of the first kind. With the help
of this relation one finds from (\ref{3.3}) in the limit $d\rightarrow 2$
the standard expansion of the logarithmic potential \cite{E53}.  The
perfect-screening rule (\ref{3.7}) becomes upon taking the limit
$d\rightarrow 2$:
\begin{eqnarray}
\sum_{\sigma_3} n_{\sigma_3}\, e_{\sigma_3} \int d{\bf r}_{3}\; 
h_{\sigma_1\sigma_2\sigma_3}^{(3)}({\bf r}_1,{\bf r}_2,{\bf r}_3)\;  r_{13}^\ell\;
T_{\ell}(\cos\theta)=\nonumber\\
=-e_{\sigma_2}\; r_{12}^\ell 
h_{\sigma_1\sigma_2}^{(2)}({\bf r}_1,{\bf r}_2) \quad (\ell=1,2,\ldots)
\label{3.9}
\end{eqnarray}
which for the one-component case corroborates a previous result \cite{V87}.

From the above results a consistency relation can be obtained. On one hand,
we can prove from (\ref{3.2}) with (\ref{3.4}) and (\ref{3.5}), upon
multiplying by ${\bf r}_{12}$ and integrating over ${\bf r}_2$: 
\begin{eqnarray}
\fl \beta\;  e_{\sigma_1}\sum_{\sigma_3}n_{\sigma_3}\;
e_{\sigma_3}\int d{\bf r}_2 \; d{\bf r}_3\; 
h^{(3)}_{\sigma_1\sigma_2\sigma_3}({\bf r}_1,{\bf r}_2,{\bf r}_3)\; 
{\bf r}_{12}\cdot\frac{\partial\phi(r_{13})}{\partial {\bf r}_1}=\nonumber\\
 = -\half\beta\; e_{\sigma_1}\sum_{\sigma_3} n_{\sigma_3}\; e_{\sigma_3}\int
d{\bf r}_3 \; h^{(2)}_{\sigma_2\sigma_3}({\bf r}_2,{\bf r}_3)\; r_{23}^2
 \nonumber\\
 +(d-2)\; \beta\; e_{\sigma_1}\, e_{\sigma_2}\int d{\bf r}_2\; 
h^{(2)}_{\sigma_1\sigma_2}({\bf r}_1,{\bf r}_2)\;\phi(r_{12}) \nonumber\\
 +\left[ d- (d-2)\; c_d \; \beta\; e_{\sigma_1}\; e_{\sigma_2}\right] 
\int d{\bf r}_2 \; h^{(2)}_{\sigma_1\sigma_2}({\bf r}_1,{\bf r}_2)
\label{3.10}
\end{eqnarray}
On the other hand, from (\ref{3.7}) for $\ell=1$ one gets after
multiplication by $r^{-d}_{12}$ and integration over ${\bf r}_2$:
\begin{eqnarray}
\fl \sum_{\sigma_3}n_{\sigma_3}\;e_{\sigma_3} 
\int d{\bf r}_2 \; d{\bf r}_3\; 
h^{(3)}_{\sigma_1\sigma_2\sigma_3}({\bf r}_1,{\bf r}_2,{\bf r}_3)\;
{\bf r}_{13}\cdot\frac{\partial\phi(r_{12})}{\partial{\bf r}_1}=\nonumber\\
\fl =(d-2)\; e_{\sigma_2}\int d{\bf r}_2\;
h^{(2)}_{\sigma_1\sigma_2}({\bf r}_1,{\bf r}_2)\;\phi(r_{12})
-(d-2)\; c_d\; e_{\sigma_2} \int d{\bf r}_2\; 
h^{(2)}_{\sigma_1\sigma_2}({\bf r}_1,{\bf r}_2)
\label{3.11}
\end{eqnarray}
Comparison of (\ref{3.10}) and (\ref{3.11}) yields an identity, which by
means of (\ref{3.5}) gets the simple form:
\begin{equation}
\sum_{\sigma_1,\sigma_2} n_{\sigma_1}\;n_{\sigma_2}\;
e_{\sigma_1}\;e_{\sigma_2}\int d{\bf r}_2\; 
h^{(2)}_{\sigma_1\sigma_2}({\bf r}_1,{\bf r}_2)\;
r^2_{12}=-\frac{2\; d}{\beta}
\label{3.12}
\end{equation}
For $d=3$ this identity reduces to the well-known sum rule that
was first obtained by Stillinger and Lovett \cite{SL68} and discussed
subsequently extensively \cite{SvW87,V85,vBF79,HS77,MMSG77,MG83}. For the
one-component case with $d>3$ its form has been found before \cite{A88}. 

The sum rule (\ref{3.12}) is independent of $c_d$, as it should be, since the
correlation functions cannot depend on the choice of an additive constant
in the potential. For $d>2$ the intermediate steps in deriving (\ref{3.12})
simplify for the choice $c_d=0$, but that is not essential for the proof. To
treat the limit $d\rightarrow 2$ one has to choose the specific value
(\ref{2.3}) for $c_d$, so that $\phi(r)$ stays finite. With that
particular choice the proof of (\ref{3.12}) remains valid in the limit $d
\rightarrow 2$. The form of (\ref{3.12}) for $d\rightarrow 2$ is consistent
with that found previously by taking $d=2$ from the start
\cite{V87,MG83}. The above derivation shows how the general form of the
Stillinger-Lovett relation for an ionic mixture reads for arbitrary $d\geq
2$. 

\section{Equilibrium ensemble and thermodynamics}
\setcounter{equation}{0} 

To prepare the ground for the derivation of additional sum rules for the
pair correlation functions of the ionic mixture we need to specify the
equilibrium ensemble for the system. A convenient choice, which has been
discussed before \cite{SvW87}, is the restricted grand-canonical
ensemble. It is a grand-canonical ensemble with particle numbers satisfying
the constraint $\sum_{\sigma} e_\sigma\; N_\sigma=q_v \; V$.  Its partition
function $Z$ depends on the volume $V$, the inverse temperature $\beta$,
the background charge density $q_v$, and $s-1$ chemical potentials
$\tilde{\mu}_\sigma (\sigma\neq 1)$. In the limit of an infinite system the
partition function leads to a thermodynamic function $\tilde{p}$ that is
defined by writing:
\begin{equation}
\lim_{V\rightarrow \infty}\;\frac{1}{V}\;
\log Z(\beta,\{\beta\tilde{\mu}_\sigma\},q_v,V)=
\beta\tilde{p}(\beta,\{\beta\tilde{\mu}_\sigma\},q_v)
\label{4.1}
\end{equation}
The energy density $u_v$ and the particle densities $n_\sigma$ for
$\sigma\neq 1$ follow by taking derivatives:
\begin{equation}
u_v=-\frac{\partial\beta\tilde{p}}{\partial\beta} \quad , \quad
n_\sigma=\frac{\partial\beta\tilde{p}}{\partial\beta\tilde{\mu}_\sigma}
\quad (\sigma = 2, \ldots , s)
\label{4.2}
\end{equation}
In writing a partial derivative with respect to one of the variables
$\beta,\{\beta\tilde{\mu}_\sigma\},q_v$, the other variables that are meant to
remain constant are suppressed. The pressure follows from $\tilde{p}$
through the relation 
\begin{equation}
p=\tilde{p}-q_v\;\frac{\partial\tilde{p}}{\partial q_v}
\label{4.3}
\end{equation}
as is proved in appendix A.

For $d>2$ a scaling argument can be used to relate the partial derivatives
of $\tilde{p}$. In fact, the potential energy satisfies the identity
\begin{eqnarray}
\fl U({\bf r}^{N_1},\ldots,{\bf r}^{N_s},V)=\lambda^{d-2}\; 
U(\lambda\; {\bf r}^{N_1},\ldots,\lambda\; {\bf r}^{N_s},\lambda^d\; V)
+\half \;\left(\lambda^{d-2}-1\right)\; c_d \sum_{\sigma} N_\sigma\;
e^2_\sigma
\nonumber\\
\label{4.4}
\end{eqnarray}
for arbitrary positive $\lambda$.  This property implies a specific scaling
behaviour of the partition function $Z$ and the thermodynamic function
$\tilde{p}$. As a consequence, the pressure and the energy density of the
ionic mixture are related as
\begin{equation}
p=\frac{d-2}{d}\; u_v -\half(d-4)\; \frac{n}{\beta}+\frac{d-2}{2d}\;
c_d\sum_\sigma n_\sigma\; e^2_\sigma
\label{4.5}
\end{equation}
with $n=\sum_\sigma n_\sigma$ the total particle density. 

The partition function $Z$, and hence $\tilde{p}$, depends on the additive
constant $c_d$ via the Hamiltonian. However, the combination (\ref{4.3}),
which gives the pressure $p$, is invariant when $c_d$ is modified. On the
other hand, the energy density $u_v$ as given by (\ref{4.2}) does depend on
$c_d$. Its dependence is such that $u_v+\half\; c_d\sum_\sigma n_\sigma \;
e^2_\sigma$ is invariant, so that (\ref{4.5}) can be satisfied. The
specific amount by which the Hamiltonian is shifted when a different choice
for $c_d$ is made depends on the particle numbers $N_\sigma$, as
(\ref{2.5}) shows. Hence, the chemical potentials $\tilde{\mu}_\sigma$
(with $\sigma\neq 1$) change as well when a different value for $c_d$ is
chosen. However, the combination $\tilde{\mu}_\sigma+\half \; c_d\;
e_\sigma \;(e_\sigma-e_1)$ is found to be invariant. Of course, the partial
densities $n_\sigma$ do not depend on $c_d$.

It should be noted that both the pressure and the energy density can be
written as a sum of a kinetic and a potential part:
\begin{equation}
p= \frac{n}{\beta}+p^{pot}\quad , \quad u_v=\frac{d\;
  n}{2\;\beta}+u^{pot}_v
\label{4.6}
\end{equation}
According to (\ref{4.5}) the potential parts of the pressure and the energy
density are related as $p^{pot}=[(d-2)/d]\; \left(u^{pot}_v+\half\;
c_d\sum_\sigma n_\sigma \; e^2_\sigma\right)$. In appendix A it is shown
how several auxiliary relations can be derived from (\ref{4.5}).

For dimension $d>2$ one may take $c_d=0$, so that (\ref{4.5}) gets a
simpler form \cite{J84}. In contrast, for $d\rightarrow 2$ one should
choose $c_d$ according to (\ref{2.3}). With that choice the energy density
$u_v$ stays finite for $d\rightarrow 2$. Hence, it drops out from
(\ref{4.5}) in the limit. As a consequence, we are left with the equation
of state for the two-dimensional ionic mixture:
\begin{equation}
p=\frac{n}{\beta}- \frac{1}{8\pi}\; \sum_\sigma   n_\sigma\; e_\sigma^2
\label{4.7}
\end{equation}
which can also be obtained directly by applying a scaling argument to a system
with a logarithmic potential \cite{SP63,M67}. The present derivation shows
how the second term at the right-hand side comes about as a consequence of
the shift $c_d$ in the potential. Incidentally, we remark that
it is essential to choose the right value for $c_d$ before taking the limit
$d\rightarrow 2$. For instance, choosing $c_d=0$ in (\ref{4.5}) and taking
the limit naively, without realizing that $u_v$ diverges in that case, 
would have resulted in an incorrect equation of state.

In closing this section we remark that alternatively one may choose to
describe the equilibrium ionic mixture by means of a full grand-canonical
ensemble with a background with fixed density \cite{M88,J84,vBF79,LL72,S89}.

\section{Thermodynamic sum rules for pair correlation functions: zeroth-
and second-moment rules} 
\setcounter{equation}{0} 

In the restricted grand-canonical ensemble the derivative of the partial
density $n_{\sigma_1}$ with respect to the chemical potential combination
$\beta\tilde{\mu}_{\sigma_2}$ is given by
\begin{equation}
\frac{Dn_{\sigma_1}}{D\beta\tilde{\mu}_{\sigma_2}}=
\frac{1}{V}\;\langle N_{\sigma_1}\; N_{\sigma_2}\rangle
-\frac{1}{V}\; \langle N_{\sigma_1}\rangle\;\langle
N_{\sigma_2}\rangle
\label{5.1}
\end{equation}
with the operator $D/D\beta\tilde{\mu}_\sigma$ defined as 
\begin{equation}
\frac{D}{D\beta\tilde{\mu}_\sigma}=\left(1-\delta_{\sigma 1}\right)\; 
\frac{\partial}{\partial\beta\tilde{\mu}_\sigma}
-\delta_{\sigma 1}\sum_{\sigma'\neq 1}\frac{e_{\sigma'}}{e_1}\; 
\frac{\partial}{\partial\beta\tilde{\mu}_{\sigma'}}
\label{5.2}
\end{equation}
The right-hand side of (\ref{5.1}) can be expressed as an integral over
the pair correlation function. As a result one finds:
\begin{equation}
n_{\sigma_1}\; n_{\sigma_2} \int d{\bf r}_2\;
h^{(2)}_{\sigma_1\sigma_2}({\bf r}_1,{\bf r}_2) =
\frac{Dn_{\sigma_1}}{D\beta\tilde{\mu}_{\sigma_2}}
-n_{\sigma_1}\;\delta_{\sigma_1\sigma_2}
\label{5.3}
\end{equation}
Upon summation over $\sigma_2$, with the weights $e_{\sigma_2}$, one
recovers the perfect-screening rule (\ref{3.5}). Taking an unweighted sum
over $\sigma_2$ and using (\ref{A.8}) to eliminate the derivative of the
particle density $n$ we find the equality:
\begin{eqnarray}
\fl \half d\; (d-4)\; n_{\sigma_1}\sum_{\sigma_2}n_{\sigma_2}
\int d{\bf r}_2\; h^{(2)}_{\sigma_1\sigma_2}({\bf r}_1,{\bf r}_2)
=-(d-2)\;\beta\;\frac{\partial n_{\sigma_1}}{\partial\beta}
+d\; q_v\;\frac{\partial n_{\sigma_1}}{\partial q_v}\nonumber\\
-\half d\; (d-2)\; n_{\sigma_1}
+\half\; (d-2)\; c_d\; \beta\; \frac{D}{D\beta\tilde{\mu}_{\sigma_1}}
\left(\sum_{\sigma_2} n_{\sigma_2}\; e_{\sigma_2}^2\right)
\label{5.4}
\end{eqnarray}
This zeroth-moment sum rule is independent of the perfect-screening
sum rule. Like that rule it is valid for each species $\sigma_1$
separately. If an unweighted sum over $\sigma_1$ is carried out, one arrives
at a less strong sum rule of the form:
\begin{eqnarray}
\fl \half\; d\;(d-4)\sum_{\sigma_1,\sigma_2} n_{\sigma_1}\;n_{\sigma_2} 
\int d{\bf r}_2\; h^{(2)}_{\sigma_1\sigma_2}({\bf r}_1,{\bf r}_2)=
-(d-2)\;\beta \;\frac{\partial n}{\partial\beta}
+d\; q_v\;\frac{\partial n}{\partial q_v}\nonumber\\
-\half d\; (d-2)\; n +\half\; (d-2)\; c_d\; \beta\; 
\sum_{\sigma_1}e^2_{\sigma_1}\; \frac{Dn}{D\beta\tilde{\mu}_{\sigma_1}}
\label{5.5}
\end{eqnarray}
For any $d>2$ one may choose $c_d=0$ in (\ref{5.4}) and (\ref{5.5}). For
$d=4$ the integrals in (\ref{5.4}) and (\ref{5.5}) drop out; the resulting
equalities are trivial consequences of the relation (\ref{A.8}). The case
$d=2$ deserves special attention, and will be discussed at the end of this
section.

The derivative of the partial density $n_{\sigma}$ with respect to the
inverse temperature $\beta$ reads 
\begin{equation}
\frac{\partial n_{\sigma}}{\partial \beta}=
-\frac{1}{V}\; \langle N_{\sigma}\; H\rangle
+\frac{1}{V}\; \langle N_{\sigma}\rangle\;\langle H\rangle
\label{5.6}
\end{equation}
Like the derivative with respect to the chemical potentials discussed
above, it can be written  in terms of integrals over Ursell functions, as
shown in appendix B:
\begin{eqnarray}
\fl (d-2)\; \beta\; \frac{\partial n_{\sigma_1}}{\partial\beta}=
-\half\; \beta\; q_v\; n_{\sigma_1}\sum_{\sigma_2}
n_{\sigma_2}\; e_{\sigma_2}\int d{\bf r}_2\; 
h^{(2)}_{\sigma_1\sigma_2}({\bf r}_1,{\bf r}_2)\; r_{12}^2\nonumber\\
 -\half\; d\; (d-4)\; n_{\sigma_1}\sum_{\sigma_2} n_{\sigma_2}\int 
d{\bf r}_2\; h^{(2)}_{\sigma_1\sigma_2}({\bf r}_1,{\bf r}_2)
-\half\; d\; (d-2)\; n_{\sigma_1}\nonumber\\
 +\half\; (d-2)\; c_d \; \beta\; n_{\sigma_1}
\left[\sum_{\sigma_2}
n_{\sigma_2}\; e^2_{\sigma_2}\int d{\bf r}_2\; 
h^{(2)}_{\sigma_1\sigma_2}({\bf r}_1,{\bf r}_2)
+e^2_{\sigma_1}\right]
\label{5.7}
\end{eqnarray}
An essential role in the proof of this identity is played by the symmetry
properties of the Ursell functions, as is discussed in appendix
C. Employing (\ref{5.3}) and (\ref{5.4}) for two of the integrals at the
right-hand side, we find that many terms cancel. In this way we obtain the
second-moment sum rule:
\begin{equation}
n_{\sigma_1}\sum_{\sigma_2}n_{\sigma_2}\; e_{\sigma_2}\int d{\bf r}_2\;
h^{(2)}_{\sigma_1\sigma_2}({\bf r}_1,{\bf r}_2)\; r_{12}^2=
-\frac{2\; d}{\beta}\;\frac{\partial n_{\sigma_1}}{\partial q_v}
\label{5.8}
\end{equation}
Summing over $\sigma_1$ with the weights $e_{\sigma_1}$ one recovers the
Stillinger-Lovett rule (\ref{3.12}). Taking the sum with equal weights
we get the second-moment identity
\begin{equation}
\sum_{\sigma_1,\sigma_2}n_{\sigma_1}\; n_{\sigma_2}\; e_{\sigma_2}\int d{\bf r}_2\;
h^{(2)}_{\sigma_1\sigma_2}({\bf r}_1,{\bf r}_2)\; r_{12}^2=
-\frac{2\; d}{\beta}\;\frac{\partial n}{\partial q_v}
\label{5.9}
\end{equation}
which is independent of the Stillinger-Lovett rule. 

The above sum rules have been derived for all $d>2$. To obtain the
corresponding rules for the case $d=2$ we choose $c_d$ according to
(\ref{2.3}) and take the limit $d \rightarrow 2$. The zeroth-order sum rule
(\ref{5.3}) retains the same form, whereas the sum rules (\ref{5.4}) and
(\ref{5.5}) become
\begin{equation}
\fl n_{\sigma_1}\sum_{\sigma_2}n_{\sigma_2}
\int d{\bf r}_2\; h^{(2)}_{\sigma_1\sigma_2}({\bf r}_1,{\bf r}_2)=
-q_v\;\frac{\partial n_{\sigma_1}}{\partial q_v}
+\frac{\beta}{8\pi}\; \frac{D}{D\beta\tilde{\mu}_{\sigma_1}}
\left(\sum_{\sigma_2} n_{\sigma_2}\; e_{\sigma_2}^2\right)
\label{5.10}
\end{equation}
and
\begin{equation}
\fl \sum_{\sigma_1,\sigma_2} n_{\sigma_1}\;n_{\sigma_2} 
\int d{\bf r}_2\; h^{(2)}_{\sigma_1\sigma_2}({\bf r}_1,{\bf r}_2)=
- q_v\;\frac{\partial n}{\partial q_v}
+\frac{\beta}{8\pi}\sum_{\sigma_1}e^2_{\sigma_1}\; 
\frac{Dn}{D\beta\tilde{\mu}_{\sigma_1}}
\label{5.11}
\end{equation}
As in the previous section, incorrect results would have been obtained from
(\ref{5.4}) and (\ref{5.5}) when the choice $c_d=0$ had been made before
evaluating the limit $d\rightarrow 2$. In contrast, the sum rules (\ref{5.8})
and (\ref{5.9}) are independent of the choice of $c_d$, so that the proof
of their validity for $d=2$ is straightforward. It may be noted that in
deriving the limiting form of the auxiliary relation (\ref{5.7}) it is
important once again to choose $c_d$ correctly before taking the limit.

The above derivation of (\ref{5.8}) for general $d$ shows how one can
combine perfect screening, symmetry and thermodynamics with the statistical
relation (\ref{5.6}) to establish a second-moment sum rule. For the special
case $d=2$ the last mentioned ingredient is not necessary, as is shown in
detail in appendix C. This particular feature of the second-moment sum rule
(\ref{5.8}) for $d=2$ has been discovered recently \cite{JS08}.

\section{Thermodynamic sum rules for pair correlation functions: 
fourth-moment rule} 
\setcounter{equation}{0}

To derive an equality for the fourth moment of the two-particle Ursell
function we start from an expression for its derivative with respect to the
inverse temperature:
\begin{eqnarray}
\fl  (d-2)\; \beta\;\frac{\partial}{\partial\beta}\left[ n_{\sigma_1}\; n_{\sigma_2}\; 
h^{(2)}_{\sigma_1\sigma_2}({\bf r}_1,{\bf r}_2)\right]=\nonumber\\
\fl =-\half\;\beta\; q_v\; n_{\sigma_1}\; n_{\sigma_2}
\sum_{\sigma_3}n_{\sigma_3}\;e_{\sigma_3} \int d{\bf r}_3\;
h^{(3)}_{\sigma_1\sigma_2\sigma_3}({\bf r}_1,{\bf r}_2,{\bf r}_3)\;
r^2_{23}\nonumber\\
\fl -\half\; d\; (d-4)\; n_{\sigma_1}\; n_{\sigma_2}
\sum_{\sigma_3}n_{\sigma_3}\int d{\bf r}_3\;
h^{(3)}_{\sigma_1\sigma_2\sigma_3}({\bf r}_1,{\bf r}_2,{\bf
  r}_3)-n_{\sigma_1}\; n_{\sigma_2}\; 
{\bf r}_{12}\cdot\frac{\partial}{\partial{\bf r}_1}\; 
h^{(2)}_{\sigma_1\sigma_2}({\bf r}_1,{\bf r}_2)\nonumber\\
\fl  -\half\; \beta\; q_v\; n_{\sigma_1}\; n_{\sigma_2}\; e_{\sigma_1}\; 
h^{(2)}_{\sigma_1\sigma_2}({\bf r}_1,{\bf r}_2)\; r^2_{12}
-d\; (d-2)\; n_{\sigma_1}\; n_{\sigma_2}\; 
h^{(2)}_{\sigma_1\sigma_2}({\bf r}_1,{\bf r}_2)\nonumber\\
\fl +\half\; (d-2)\; c_d\; \beta\; n_{\sigma_1}\; n_{\sigma_2}\;
\left[\sum_{\sigma_3} n_{\sigma_3}\; e^2_{\sigma_3}\int d{\bf r}_3 \;
h^{(3)}_{\sigma_1\sigma_2\sigma_3}({\bf r}_1,{\bf r}_2,{\bf r}_3)\right.\nonumber\\
\left. +\left( e^2_{\sigma_1}+e^2_{\sigma_2}\right)\; 
h^{(2)}_{\sigma_1\sigma_2}({\bf r}_1,{\bf r}_2)\right]
\label{6.1}
\end{eqnarray}
The proof of this identity is sketched in appendix B. Multiplying both
sides with $e_{\sigma_1}\; e_{\sigma_2}\; \phi(r_{12})$, integrating over
${\bf r}_2$ and summing over $\sigma_1$ and $\sigma_2$, we get an
expression for the derivative of the potential-energy density (\ref{B.1}):
\begin{eqnarray}
\fl  2\; (d-2)\; \beta\;\frac{\partial u^{pot}_v}
{\partial\beta}=\nonumber\\
\fl =-\half\;\beta\; q_v\; 
\sum_{\sigma_1,\sigma_2,\sigma_3}
n_{\sigma_1}\; n_{\sigma_2}\; n_{\sigma_3}\;
e_{\sigma_1}\; e_{\sigma_2}\;e_{\sigma_3}
\int d{\bf r}_2\; d{\bf r}_3\;
h^{(3)}_{\sigma_1\sigma_2\sigma_3}({\bf r}_1,{\bf r}_2,{\bf r}_3)\;
r^2_{23}\; \phi(r_{12})\nonumber\\
\fl -\half\; d\; (d-4)\sum_{\sigma_1,\sigma_2,\sigma_3}
 n_{\sigma_1}\; n_{\sigma_2}\;n_{\sigma_3}
e_{\sigma_1}\; e_{\sigma_2}
\int d{\bf r}_2\; d{\bf r}_3\;
h^{(3)}_{\sigma_1\sigma_2\sigma_3}({\bf r}_1,{\bf r}_2,{\bf
  r}_3)\; \phi(r_{12})\nonumber\\
\fl -\sum_{\sigma_1,\sigma_2}
n_{\sigma_1}\; n_{\sigma_2}\; e_{\sigma_1}\; e_{\sigma_2}\; 
\int d{\bf r}_2\; {\bf r}_{12}\cdot\left[\frac{\partial}{\partial{\bf r}_1}\; 
h^{(2)}_{\sigma_1\sigma_2}({\bf r}_1,{\bf r}_2)\right]\; \phi(r_{12})\nonumber\\
\fl  -\half\; \beta\; q_v
\sum_{\sigma_1,\sigma_2} n_{\sigma_1}\; n_{\sigma_2}\; e^2_{\sigma_1}\;
e_{\sigma_2}\; \int d{\bf r}_2\; 
h^{(2)}_{\sigma_1\sigma_2}({\bf r}_1,{\bf r}_2)\; r^2_{12}\; \phi(r_{12})
-2\; d\; (d-2)\; u^{pot}_v\nonumber\\
\fl +\half\; (d-2)\; c_d\; \beta\sum_{\sigma_1,\sigma_2}
n_{\sigma_1}\; n_{\sigma_2}\; e_{\sigma_1}\; e_{\sigma_2}\; 
\left[ \sum_{\sigma_3} n_{\sigma_3}\; e^2_{\sigma_3}\; 
\int d{\bf r}_2\; d{\bf r}_3\; 
h^{(3)}_{\sigma_1\sigma_2\sigma_3}({\bf r}_1,{\bf r}_2,{\bf r}_3)\; 
\phi(r_{12})\right. \nonumber\\
\left. +\left(e^2_{\sigma_1}+e^2_{\sigma_2}\right)
\int d{\bf r}_2\; h^{(2)}_{\sigma_1\sigma_2}({\bf r}_1,{\bf r}_2)\; \phi(r_{12})\right]
\label{6.2}
\end{eqnarray}
The first term at the right-hand side can be expressed in moments of the
two-particle Ursell functions by using the relation (\ref{C.4}), which 
follows from the symmetry properties of the three-particle Ursell
function. Likewise, the second and the sixth terms can be
rewritten by means of the symmetry relation (\ref{C.2}). In the third term we can
carry out a partial integration and use the identity
\begin{equation}
{\bf r}_{12}\cdot\frac{\partial \phi(r_{12})}{\partial{\bf r}_1}=
-(d-2)\; \phi(r_{12})+(d-2)\; c_d
\label{6.3}
\end{equation}
As a result we arrive at a relation involving the
zeroth, the second and the fourth moments of the two-particle Ursell
function:
\begin{eqnarray}
\fl d\; \beta^2\; q_v^2\sum_{\sigma_1,\sigma_2}
n_{\sigma_1}\; n_{\sigma_2}\; e_{\sigma_1}\;e_{\sigma_2}\int d{\bf r}_2\;  
h^{(2)}_{\sigma_1\sigma_2}({\bf r}_1,{\bf r}_2)\; r^4_{12}\nonumber\\
\fl +2\; d\; (d-6)\; (d+2)\; \beta\; q_v
\sum_{\sigma_1,\sigma_2}
n_{\sigma_1}\; n_{\sigma_2}\; e_{\sigma_1}\int d{\bf r}_2\;  
h^{(2)}_{\sigma_1\sigma_2}({\bf r}_1,{\bf r}_2)\; r^2_{12}\nonumber\\
\fl -4\;d^2 \; (d-4)\; (d+2)
\sum_{\sigma_1,\sigma_2}
n_{\sigma_1}\; n_{\sigma_2}\int d{\bf r}_2\;  
h^{(2)}_{\sigma_1\sigma_2}({\bf r}_1,{\bf r}_2)=\nonumber\\
\fl =-8\; (d-2)^2\; (d+2)\; \beta^2\;\frac{\partial u_v^{pot}}{\partial\beta}
-16\; (d-1)\; (d-2)\; (d+2)\; \beta \; u_v^{pot}\nonumber\\
\fl +2\; (d-2)\; (d+2)\; c_d\; \beta \left[ 2\; \beta\; q_v 
\sum_{\sigma_1,\sigma_2} n_{\sigma_1}\; n_{\sigma_2}\; e^2_{\sigma_1}\;
e_{\sigma_2} \int d{\bf r}_2\; 
h^{(2)}_{\sigma_1\sigma_2}({\bf r}_1,{\bf r}_2)\; r^2_{12}\right. \nonumber\\ 
\fl \left. +d\; (d-6) \sum_{\sigma_1,\sigma_2} n_{\sigma_1}\; n_{\sigma_2}\; e^2_{\sigma_1}\;
\int d{\bf r}_2\; 
h^{(2)}_{\sigma_1\sigma_2}({\bf r}_1,{\bf r}_2)
+(d^2-6d+4)\sum_{\sigma_1} n_{\sigma_1}\; e^2_{\sigma_1}\right]\nonumber\\
\fl -2\; (d-2)^2\; (d+2)\; c_d^2\; \beta^2\left[
 \sum_{\sigma_1,\sigma_2} n_{\sigma_1}\; n_{\sigma_2}\; e^2_{\sigma_1}\;
e^2_{\sigma_2}\int d{\bf r}_2\; 
h^{(2)}_{\sigma_1\sigma_2}({\bf r}_1,{\bf r}_2)
+\sum_{\sigma_1} n_{\sigma_1}\; e^4_{\sigma_1}\right]
\label{6.4}
\end{eqnarray}
The zeroth and second moments at both sides of this relation can be
replaced by the thermodynamic expressions given in (\ref{5.3}) and
(\ref{5.8}). The ensuing derivatives with respect to the chemical
potentials may be eliminated with the help of (\ref{A.8}). Furthermore, at the
right-hand side the full energy density can be introduced with the help of
(\ref{4.6}). These manipulations lead to an expression for the fourth
moment of the Ursell function in terms of thermodynamic derivatives only:
\begin{eqnarray}
 \fl d\; \beta^2\; q_v^2 \sum_{\sigma_1,\sigma_2}
n_{\sigma_1}\; n_{\sigma_2}\; e_{\sigma_1}\;e_{\sigma_2}\int d{\bf r}_2\;  
h^{(2)}_{\sigma_1\sigma_2}({\bf r}_1,{\bf r}_2)\; r^4_{12}=\nonumber\\
\fl =-8\; (d-2)^2\; (d+2)\; \beta^2\; \frac{\partial u_v}{\partial \beta}
-16\; (d-1)\; (d-2)\; (d+2)\; \beta\; u_v\nonumber\\
\fl +4\; d\; (d-2)\; (d-4)\; (d+2)\; \beta\; \frac{\partial n}{\partial \beta}
+4\; d^2\; (d-4)\; (d+2)\; q_v\; \frac{\partial n}{\partial q_v} \nonumber\\
\fl -4\; (d-2)\; (d+2)\; c_d\; \beta \left[
(d-2)\; \beta\; \frac{\partial}{\partial\beta}
+d\; q_v\; \frac{\partial}{\partial q_v}+d-2\right]
\left(\sum_\sigma n_{\sigma}\; e^2_{\sigma} \right)
\label{6.5}
\end{eqnarray}
Upon using (\ref{A.9}) we find that the right-hand side is proportional to
the derivative of the pressure $p$ (in the form of (\ref{4.5})) with
respect to $q_v$. In this way we have found the rather elegant
fourth-moment sum rule
\begin{equation}
\fl \sum_{\sigma_1,\sigma_2}
n_{\sigma_1}\; n_{\sigma_2}\; e_{\sigma_1}\;e_{\sigma_2}\int d{\bf r}_2\;  
h^{(2)}_{\sigma_1\sigma_2}({\bf r}_1,{\bf r}_2)\; r^4_{12}
=-\frac{8d(d+2)}{\beta q_v}\; \frac{\partial p}{\partial q_v}
\label{6.6}
\end{equation}
It is a generalization to arbitrary $s$ and $d$ of the well-known
compressibility rule that has been established for the one-component plasma
in two \cite{V87} and three \cite{V85,VH75,B78,SC85} dimensions and for the
three-dimensional ionic mixture \cite{SvW87,vBF79}. Whereas the second
moments, as given by the sum rules (\ref{3.12}), (\ref{5.8}) and
(\ref{5.9}), are linear in $d$, the fourth moment turns out to be quadratic
in $d$.

For all $d>2$ we may put $c_d=0$, as before, so that the relations
(\ref{6.1})--(\ref{6.5}) become somewhat simpler. To discuss the case $d=2$
we must choose $c_d$ as in (\ref{2.3}). Upon taking the limit $d
\rightarrow 2$ the terms in (\ref{6.1})--(\ref{6.5}) containing $c_d$
remain finite, so that they cannot be omitted. However, the final result
(\ref{6.6}) does not depend on $c_d$ explicitly, so that it remains valid
as such in the limit $d\rightarrow 2$. Hence, we have established the
fourth-moment rule (\ref{6.6}) for all $d\geq 2$.

As a final remark we point out that a shorter proof of the fourth-moment
rule for the special case $d=2$ can be found from a particular symmetry
relation connecting second and fourth moments, as discussed in appendix
C. The derivative $\partial h^{(2)}_{\sigma_1\sigma_2}/\partial\beta$ is
not needed in that line of reasoning.

\section{Concluding remarks}
\setcounter{equation}{0}

By making a systematic use of the properties of the restricted
grand-canonical ensemble and the hierarchy equations for the correlation
functions we have been able to derive the sum rules that govern the first
few moments of the two-particle Ursell functions for a multi-component ionic
mixture with an arbitrary spatial dimension $d\geq 2$. The dependence on
$d$ of the various moments has been determined in detail. While most
discussions in the literature had to treat two-dimensional mixtures with a
logarithmic potential as a separate case, we have shown that a unified
description of mixtures for all $d\geq 2$ is indeed  possible by making a
careful choice of additive constants in the potential.

Our main results for the moments of the two-particle Ursell function are
presented in (\ref{3.5}), (\ref{3.12}), (\ref{5.3})--(\ref{5.5}),
(\ref{5.8})--(\ref{5.9}) and (\ref{6.6}). The ensuing results for the
moments of the two-particle correlation function follow by replacing
$h^{(2)}_{\sigma_1\sigma_2}$ with
$g^{(2)}_{\sigma_1\sigma_2}-1$. Whereas the zeroth-moment
perfect-screening rules (\ref{3.5}) and the second-moment rule (\ref{3.12})
could be derived without invoking thermodynamical properties, the proof of
the other sum rules had to be based on statistical ensemble
theory. Accordingly, the ensuing rules in sections 5 and 6 depend on
thermodynamical derivatives with respect to the basic variables describing
ionic mixtures in a restricted grand-canonical ensemble, viz.\ $\beta$,
$\{\beta\tilde{\mu}_{\sigma}\}$ (for $\sigma=2,\dots,s$) and $q_v$.

If one wishes, one may express the sum rules in terms of derivatives with
respect to a different set of independent variables involving -- apart from
$\beta$ -- the chemical potentials $\{\mu_{\sigma}\}$ with
$\sigma=1,\dots,s$, in a way described previously \cite{SvW87}. For
completeness we give the sum rules (\ref{5.3}), (\ref{5.8}) and (\ref{6.6})
in terms of derivatives with respect to these alternative variables:
\begin{equation}
\fl n_{\sigma_1}\; n_{\sigma_2}    
\int d{\bf r}_2\;
h^{(2)}_{\sigma_1\sigma_2}({\bf r}_1,{\bf r}_2) =
\frac{1}{\beta}\;  \left(
\frac{\partial n_{\sigma_1}}{\partial \mu_{\sigma_2}}-
\frac{1}{S}\; \frac{\partial q_v}{\partial \mu_{\sigma_1}}\;
\frac{\partial q_v}{\partial \mu_{\sigma_2}}\right)-n_{\sigma_1}\; 
\delta_{\sigma_1\sigma_2}
\label{7.1}
\end{equation}
\begin{equation}
\fl n_{\sigma_1}\sum_{\sigma_2}n_{\sigma_2}\; e_{\sigma_2}\int d{\bf r}_2\;
h^{(2)}_{\sigma_1\sigma_2}({\bf r}_1,{\bf r}_2)\; r_{12}^2=
-\frac{2\;d}{\beta\; S}\; \frac{\partial q_v}{\partial \mu_{\sigma_1}}
\label{7.2}
\end{equation}
\begin{equation}
\fl \sum_{\sigma_1,\sigma_2}
n_{\sigma_1}\; n_{\sigma_2}\; e_{\sigma_1}\;e_{\sigma_2}\int d{\bf r}_2\;  
h^{(2)}_{\sigma_1\sigma_2}({\bf r}_1,{\bf r}_2)\; r^4_{12} =
-\frac{8\; d\; (d+2)}{\beta\; S}
\label{7.3}
\end{equation}
with the abbreviation $S=\sum_\sigma e_\sigma \partial q_v/\partial
\mu_\sigma$. As before, in writing the partial derivatives at the
right-hand sides the independent variables that are kept constant are
suppressed.

The sum rules discussed in this paper are essential in understanding the
equilibrium fluctuations in an ionic mixture. In particular, the
fluctuations in the partial densities, the pressure and the energy density
are governed by these rules, as has been shown in \cite{vWS87} for the
three-dimensional case. The fluctuation formulas in turn are necessary in
order to determine specific dynamical properties of the ionic mixture, such
as the time evolution of the collective modes \cite{B83,SS90}.

\ack I am indebted to B.\ Jancovici for having suggested this problem and
to A.J.\ van Wonderen for many stimulating discussions.

\appendix
\section{Stability, thermodynamic pressure and some auxiliary relations}
\setcounter{section}{1} 

In this appendix we shall first discuss the stability of ionic mixtures in
arbitrary dimension. Furthermore, we shall establish the relation between
the thermodynamic pressure and the partition function in the restricted
grand-canonical ensemble. Finally, a few thermodynamic auxiliary relations
will be derived.

By generalizing the argument given in \cite{LN75} so as to be applicable to
a mixture in arbitrary dimension $d>2$ one finds the bound:
\begin{equation}
\fl  U\geq -\frac{d}{2\pi (d+2)(d-2)}\; q_v^{(d-2)/d}\; 
\left[\Gamma(\half d+1)\right]^{2/d}\; \sum_{\sigma}
 N_{\sigma}\; e_\sigma^{1+2/d}
-\half \; c_d\sum_\sigma   N_\sigma\; e_\sigma^2
\label{A.1}
\end{equation}
which for the one-component case (and $c_d=0$) agrees with the bound
presented by Sari et al \cite{SM76,SMC76}.  Taking moreover $d=3$ one
recovers the result in \cite{LN75}. For $d \rightarrow 2$ and $c_d=0$ the
bound in (\ref{A.1}) goes to $-\infty$ , so that it becomes
useless. However, upon choosing $c_d$ as in (\ref{2.3}) the inequality
(\ref{A.1}) becomes in the limit $d\rightarrow 2$:
\begin{equation}
U\geq -\frac{1}{8\pi}\; \sum_\sigma N_\sigma\; e_\sigma^2\;
\log\left(\frac{\pi q_v}{e_\sigma}\right) -\half\left(c+\frac{3}{8\pi}\right)
\sum_\sigma  N_\sigma\; e_\sigma^2
\label{A.2}
\end{equation}
For the one-component case (and the choice $c=0$) this inequality has been
derived previously \cite{SM76}. It should be remarked that different bounds
have been obtained in the past \cite{M68}-\cite{C99}. For our present
discussion these are not relevant, since we only wish to confirm here
that the multi-component ionic mixture is stable for arbitrary $d\geq 2$.

Furthermore, we want to derive the relation (\ref{4.3}) between the the
pressure $p$ and the thermodynamic function $\tilde{p}$, which follows from
the partition function according to (\ref{4.1}).  Generalizing the
definition of the (thermal) pressure in a one-component plasma by Choquard
et al \cite{CFG80} to an ionic mixture, we write it as the
derivative of the free energy $F$ with respect to the volume $V$ at
constant temperature $T$, constant (average) particle numbers $n_\sigma V$
(for $\sigma=2,\dots,s$) and constant total background charge $q_v V$:
\begin{equation}
p=-\left(\frac{\partial F(T,\{ n_\sigma \},q_v,V)}{\partial V}\right)_
{T,\{ n_{\sigma} V\}, q_v V}
\label{A.3}
\end{equation}
Taking account of the implicit dependence on $V$ we get
\begin{equation}
 p =-f_v+\sum_{\sigma(\neq 1)} n_\sigma \;
\left( \frac{\partial f_v}{\partial n_\sigma}\right)_{T,\{
  n_{\sigma'}\},q_v}
+q_v \; \left(\frac{\partial f_v}{\partial q_v}\right)_{T,\{ n_\sigma \}}
\label{A.4}
\end{equation}
with $f_v(T,\{ n_\sigma \},q_v)=F/V$ the free energy density. The
construction of the restricted grand-canonical ensemble implies the
relations \cite{SvW87}:
\begin{eqnarray}
df_v =-s_v\; dT +\sum_{\sigma(\neq 1)}\tilde{\mu}_\sigma\; 
dn_\sigma +\tilde{\mu}_q \; dq_v
\label{A.5}\\
\tilde{p}= -f_v+\sum_{\sigma(\neq 1)} \tilde{\mu}_\sigma \; n_\sigma 
\label{A.6}
\end{eqnarray}
with $s_v$ the entropy density and
$\tilde{\mu}_q=-\partial\tilde{p}/\partial q_v$. Hence, (\ref{A.4}) can be
written as:
\begin{equation}
p=-f_v+\sum_{\sigma(\neq 1)}\tilde{\mu}_\sigma\; n_\sigma +\tilde{\mu}_q \;
q_v
\label{A.7}
\end{equation}
Comparison of (\ref{A.6}) and (\ref{A.7}) yields the relation between
$\tilde{p}$ and the pressure $p$ that we wished to prove.

Finally, in the main text we need several equalities involving partial
derivatives of thermodynamic quantities. Upon differentiating the relation
(\ref{4.5}) with respect to $\beta\tilde{\mu}_\sigma$, at constant $\beta$
and $q_v$, we get:
\begin{eqnarray}
 (d-2)\;\beta\;\frac{\partial n_\sigma}{\partial\beta}=d\; q_v\; 
\frac{\partial n_\sigma}{\partial q_v}-d\; n_\sigma
-\half\; d\;(d-4)\; \frac{Dn}{D\beta\tilde{\mu}_\sigma}\nonumber\\
+\half\; (d-2)\; c_d\; \beta\; 
\frac{D}{D\beta\tilde{\mu}_\sigma}\left(\sum_{\sigma'} n_{\sigma'}\; 
e_{\sigma'}^2\right)
\label{A.8}
\end{eqnarray}
with the operator $D/D\beta\tilde{\mu}_\sigma$ defined in (\ref{5.2}).
Likewise, differentiation of (\ref{4.5}) with respect to $\beta$ yields:
\begin{eqnarray}
 (d-2)\; \beta\; \frac{\partial u_v}{\partial\beta}=
d\; q_v\; \frac{\partial u_v}{\partial q_v}
-2\; (d-1)\; u_v+
\half\; d\; (d-4)\; \frac{\partial n}{\partial\beta}\nonumber\\
-\half\; (d-2)\; c_d\; 
\left(1+\beta\;\frac{\partial}{\partial\beta}\right)\;
\left(\sum_{\sigma} n_\sigma\; e_\sigma^2\right)
\label{A.9}
\end{eqnarray}

For $d>2$ we may choose $c_d=0$, so that the last terms at the right-hand
sides of (\ref{A.8}) and (\ref{A.9}) drop out. For $d\rightarrow 2$ we
choose $c_d$ as in (\ref{2.3}). When the limit is taken, the left-hand sides
of (\ref{A.8}) and (\ref{A.9}) disappear, while the last terms at the
right-hand sides yield a finite contribution. As a result we get for
$d=2$:
\begin{equation} 
q_v\; \frac{\partial n_\sigma}{\partial q_v}-n_\sigma
+\frac{D}{D\beta\tilde{\mu}_\sigma}\left(n-\frac{\beta}{8\pi}
\sum_{\sigma'} n_{\sigma'}\; e_{\sigma'}^2\right)
=0
\label{A.10}
\end{equation}
and 
\begin{equation}
  q_v\; \frac{\partial u_v}{\partial q_v}- u_v 
-\frac{\partial n}{\partial\beta}
+\frac{1}{8\pi}\;
\left(1+\beta\;\frac{\partial}{\partial\beta}\right)\;
\left(\sum_{\sigma} n_\sigma\; e_\sigma^2\right)=0
\label{A.11}
\end{equation}
The auxiliary relations (\ref{A.8}) and (\ref{A.9}) have been used in the
main text.

\section{Derivatives of densities and Ursell functions with
  respect to the inverse temperature}

In deriving the second- and fourth-moment sum rules we need expressions for
the derivatives of the partial densities and the two-particle Ursell
functions with respect to $\beta$. The derivative of $n_{\sigma}$ with
respect to $\beta$ follows by evaluating its formal expression: $\partial
n_{\sigma}/\partial \beta=-\langle N_\sigma H\rangle/V +\langle
N_\sigma\rangle \langle H\rangle/V$.  The average $\langle H\rangle$ of
the Hamiltonian is proportional to the internal energy $u_v$, which is the
sum of a kinetic and a potential part of the form (\ref{4.6}). The latter
can be written as an integral over the two-particle Ursell function:
\begin{equation}
u^{pot}_v=\half \sum_{\sigma_1,\sigma_2}n_{\sigma_1}\;n_{\sigma_2}\;
e_{\sigma_1}\;e_{\sigma_2}
\int d{\bf r}_2\; h^{(2)}_{\sigma_1\sigma_2}({\bf r}_1,{\bf r}_2) \; 
\phi(r_{12})
\label{B.1}
\end{equation}
The average $\langle N_\sigma H\rangle$ can likewise be expressed in terms
of integrals over Ursell functions. As a result we find
\begin{eqnarray}
\fl \beta\; \frac{\partial n_{\sigma_1}}{\partial\beta}=
-\half \beta\; n_{\sigma_1} \sum_{\sigma_2,\sigma_3}
n_{\sigma_2}\; n_{\sigma_3}\; e_{\sigma_2}\; e_{\sigma_3}
\int d{\bf r}_2\; d{\bf r}_3\; 
h^{(3)}_{\sigma_1\sigma_2\sigma_3}({\bf r}_1,{\bf r}_2,{\bf r}_3)\; 
\phi(r_{23})\nonumber\\
-\beta\; n_{\sigma_1}\; e_{\sigma_1}\sum_{\sigma_2}
n_{\sigma_2}\; e_{\sigma_2}\int d{\bf r}_2 \; 
h^{(2)}_{\sigma_1\sigma_2}({\bf r}_1,{\bf r}_2)\; \phi(r_{12})\nonumber\\
-\half d\; n_{\sigma_1}\sum_{\sigma_2} n_{\sigma_2}\int d{\bf r}_2\;
h^{(2)}_{\sigma_1\sigma_2}({\bf r}_1,{\bf r}_2)-\half \; d\; n_{\sigma_1}
\label{B.2}
\end{eqnarray}
Employing the symmetry relation (\ref{C.2}) to eliminate the integral with
the three-particle Ursell function, we arrive at (\ref{5.7}).

Furthermore, we need an expression for the derivative of the two-particle
Ursell function with respect to $\beta$. In the restricted grand-canonical
ensemble one has quite generally $\partial \langle f
\rangle/\partial\beta=-\langle f\; H\rangle+\langle f\rangle\; \langle H
\rangle$ for an arbitrary phase function $f$. Taking
$f=\sum_{\alpha_1,\alpha_2}\!\!\!\!\!\!\!\!\!\!  ' \;\;\;\; \delta({\bf
r}_1-{\bf r}_{\sigma_1\alpha_1})\;\delta({\bf r}_2-{\bf
r}_{\sigma_2\alpha_2})$ and using (\ref{2.5}) one derives
\begin{eqnarray}
\fl \beta\frac{\partial}{\partial\beta}\left[ n_{\sigma_1}\; n_{\sigma_2}\; 
h^{(2)}_{\sigma_1\sigma_2}({\bf r}_1,{\bf r}_2)\right]=\nonumber\\
\fl =-\half\; \beta\; n_{\sigma_1}\; n_{\sigma_2}\sum_{\sigma_3,\sigma_4}
n_{\sigma_3}\; n_{\sigma_4}\; e_{\sigma_3}\; e_{\sigma_4}
\int d{\bf r}_3\; d{\bf r}_4\; 
h^{(4)}_{\sigma_1\sigma_2\sigma_3\sigma_4}
({\bf r}_1,{\bf r}_2,{\bf r}_3,{\bf r}_4)\; \phi(r_{34})\nonumber\\
\fl -\half\; d\;  n_{\sigma_1}\; n_{\sigma_2} \sum_{\sigma_3}  n_{\sigma_3}
\int d{\bf r}_3\; h^{(3)}_{\sigma_1\sigma_2\sigma_3}
({\bf r}_1,{\bf r}_2,{\bf r}_3)\nonumber\\
\fl -\beta\; n_{\sigma_1}\; n_{\sigma_2} \sum_{\sigma_3}  n_{\sigma_3}\;
e_{\sigma_3} \int d{\bf r}_3\; 
h^{(3)}_{\sigma_1\sigma_2\sigma_3}({\bf r}_1,{\bf r}_2,{\bf r}_3)
\; \left[e_{\sigma_1}\; \phi(r_{13})+e_{\sigma_2}\;
  \phi(r_{23})\right]\nonumber\\
\fl -d\; n_{\sigma_1}\; n_{\sigma_2}\; 
h^{(2)}_{\sigma_1\sigma_2}({\bf r}_1,{\bf r}_2)
-\beta\; n_{\sigma_1}\; n_{\sigma_2}\; e_{\sigma_1}\; e_{\sigma_2}\; 
h^{(2)}_{\sigma_1\sigma_2}({\bf r}_1,{\bf r}_2)\; \phi(r_{12})\nonumber\\
\fl -\half\; \beta\;  n_{\sigma_1}\; n_{\sigma_2}
\sum_{\sigma_3,\sigma_4}
n_{\sigma_3}\; n_{\sigma_4}\; e_{\sigma_3}\; e_{\sigma_4}
\int d{\bf r}_3\; d{\bf r}_4\; 
\left[h^{(2)}_{\sigma_1\sigma_3}({\bf r}_1,{\bf r}_3)\;
h^{(2)}_{\sigma_2\sigma_4}({\bf r}_2,{\bf r}_4)\right. \nonumber\\
\left. + h^{(2)}_{\sigma_1\sigma_4}({\bf r}_1,{\bf r}_4)\; 
h^{(2)}_{\sigma_2\sigma_3}({\bf r}_2,{\bf r}_3)\right] \;
\phi(r_{34})\nonumber\\
\fl -\beta \; n_{\sigma_1}\; n_{\sigma_2}
\sum_{\sigma_3} n_{\sigma_3}\; e_{\sigma_3}\int d{\bf r}_3\; 
\left[e_{\sigma_1}\; h^{(2)}_{\sigma_2\sigma_3}({\bf r}_2,{\bf r}_3) \; 
\phi(r_{13})+
e_{\sigma_2}\; h^{(2)}_{\sigma_1\sigma_3}({\bf r}_1,{\bf r}_3) \; 
\phi(r_{23})\right]\nonumber\\
-\beta\; n_{\sigma_1}\; n_{\sigma_2} \; e_{\sigma_1}\; e_{\sigma_2}\; \phi(r_{12})
\label{B.3}
\end{eqnarray}
For large separation of the position arguments the left-hand side vanishes
faster than any inverse power of $r_{12}$. At the right-hand side, the
first three integrals and the two terms proportional to
$h^{(2)}_{\sigma_1\sigma_2}$ share this feature. However, the property of
being of short range is not obviously true for the last two integral terms,
while it is certainly false for the final term, which is proportional to
$\phi(r_{12})$ and hence of long range. Nevertheless, by employing
(\ref{3.4}) and (\ref{3.5}) one may rewrite the sum of these terms in a
form that shows their short-range character as a function of $r_{12}$
explicitly:
\begin{eqnarray}
\fl -\frac{\beta}{d-2}\; n_{\sigma_1}\; n_{\sigma_2}\sum_{\sigma_3,\sigma_4}
n_{\sigma_3}\;n_{\sigma_4}\;e_{\sigma_3}\;e_{\sigma_4}
\int d{\bf r}_3\; h^{(2)}_{\sigma_1\sigma_3}({\bf r}_1,{\bf r}_3)\;
{\bf r}_{13}\cdot\frac{\partial \phi(r_{23})}{\partial{\bf r}_2}\nonumber\\
\times \int_{r_{24}>r_{23}} d{\bf r}_4\; 
h^{(2)}_{\sigma_2\sigma_4}({\bf r}_2,{\bf r}_4)\nonumber\\
\fl +\beta\; n_{\sigma_1}\; n_{\sigma_2} \sum_{\sigma_3,\sigma_4}
n_{\sigma_3}\;n_{\sigma_4}\;e_{\sigma_3}\;e_{\sigma_4}
\int d{\bf r}_3\; h^{(2)}_{\sigma_2\sigma_3}({\bf r}_2,{\bf r}_3)\;
\left[\phi(r_{13})-c_d\right]\nonumber\\
\times\int_{r_{14}>r_{13}} d{\bf r}_4\; 
h^{(2)}_{\sigma_1\sigma_4}({\bf r}_1,{\bf r}_4)\nonumber\\
\fl +\beta\; n_{\sigma_1}\; n_{\sigma_2}\; e_{\sigma_2} 
\; \left[\phi(r_{12})-c_d\right]\sum_{\sigma_3}
n_{\sigma_3}\; e_{\sigma_3}\; \int_{r_{13}>r_{12}}d{\bf
  r}_3\;h^{(2)}_{\sigma_1\sigma_3}({\bf r}_1,{\bf r}_3)
\label{B.4}
\end{eqnarray}
Substituting these terms and using moreover the symmetry relation
(\ref{C.6}) in the first term at the right-hand side of (\ref{B.3}), we
arrive at the somewhat simpler expression (\ref{6.1}) given in the main
text. It should be noted that at the right-hand side of (\ref{6.1}) the
potential does not occur explicitly any more.

\section{Symmetry relations}

The Ursell functions are symmetric under a permutation of both their
position arguments ${\bf r}_i$ and their component labels $\sigma_i$. From
that symmetry one proves
\begin{eqnarray}
\fl (d-2)\sum_{\sigma_2,\sigma_3}n_{\sigma_2}\;n_{\sigma_3}\;e_{\sigma_2}\;e_{\sigma_3}
\int d{\bf r}_2\; d{\bf r}_3\; 
h^{(3)}_{\sigma_1\sigma_2\sigma_3}({\bf r}_1,{\bf r}_2,{\bf r}_3)\; 
\left[\phi(r_{23})-c_d\right]  =\nonumber\\
\fl =2\sum_{\sigma_2,\sigma_3}n_{\sigma_2}\;n_{\sigma_3}\;e_{\sigma_2}\;e_{\sigma_3}
\int d{\bf r}_2\; d{\bf r}_3\; 
h^{(3)}_{\sigma_1\sigma_2\sigma_3}({\bf r}_1,{\bf r}_2,{\bf r}_3)\; 
{\bf r}_{12}\cdot\frac{\partial \phi(r_{23})}{\partial {\bf r}_2}
\label{C.1}
\end{eqnarray}
At the right-hand side we use (\ref{3.10}). Employing moreover the
perfect-screening relations (\ref{3.5}) and (\ref{3.6}) we get the identity:
\begin{eqnarray}
\fl (d-2)\; \beta \sum_{\sigma_2,\sigma_3}n_{\sigma_2}\; n_{\sigma_3}\; 
e_{\sigma_2}\; e_{\sigma_3}\int d{\bf r}_2\; d{\bf r}_3\;  
h^{(3)}_{\sigma_1\sigma_2\sigma_3}({\bf r}_1,{\bf r}_2,{\bf r}_3)\; 
\phi(r_{23})\nonumber\\
\fl +2\; (d-2)\; \beta \; e_{\sigma_1}\; \sum_{\sigma_2}n_{\sigma_2}\; e_{\sigma_2}
\int d{\bf r}_2\; h^{(2)}_{\sigma_1\sigma_2}({\bf r}_1,{\bf r}_2) \; 
\phi(r_{12})=\nonumber\\
\fl =\beta\; q_v\sum_{\sigma_2}n_{\sigma_2}\; e_{\sigma_2}\int d{\bf r}_2\; 
h^{(2)}_{\sigma_1\sigma_2}({\bf r}_1,{\bf r}_2) \; r_{12}^2
-2\; d\sum_{\sigma_2}n_{\sigma_2}\int d{\bf r}_2\; 
h^{(2)}_{\sigma_1\sigma_2}({\bf r}_1,{\bf r}_2)\nonumber\\
-(d-2)\; c_d\; \beta\; \left[\sum_{\sigma_2}n_{\sigma_2}\;
  e_{\sigma_2}^2 \int d{\bf r}_2\; h^{(2)}_{\sigma_1\sigma_2}({\bf r}_1,{\bf r}_2)
+  e_{\sigma_1}^2\right]
\label{C.2}
\end{eqnarray}
which is used in section 6 and appendix B.

A second identity is obtained by starting from an equality that is
analogous to (\ref{C.1}) and follows likewise from the symmetry of the
three-particle Ursell function:
\begin{eqnarray}
\fl (d-2)\sum_{\sigma_2,\sigma_3}n_{\sigma_2}\;n_{\sigma_3}\;e_{\sigma_2}\;e_{\sigma_3}
\int d{\bf r}_2\; d{\bf r}_3\; 
h^{(3)}_{\sigma_1\sigma_2\sigma_3}({\bf r}_1,{\bf r}_2,{\bf r}_3)\; 
\left[\phi(r_{23})-c_d\right]\nonumber\\
\times \left[r^2_{12}+{\bf r}_{12}\cdot{\bf r}_{23}
+2\; r^{-2}_{23}\;({\bf r}_{12}\cdot{\bf r}_{23})^2\right]=\nonumber\\
\fl =2\sum_{\sigma_2,\sigma_3}n_{\sigma_2}\;n_{\sigma_3}\;e_{\sigma_2}\;e_{\sigma_3}
\int d{\bf r}_2\; d{\bf r}_3\; 
h^{(3)}_{\sigma_1\sigma_2\sigma_3}({\bf r}_1,{\bf r}_2,{\bf r}_3)\; r^2_{12}\;
{\bf r}_{12}\cdot\frac{\partial \phi(r_{23})}{\partial {\bf r}_{2}}
\label{C.3}
\end{eqnarray}
The right-hand side can be expressed in terms of two-particle Ursell
functions by employing the hierarchy equation (\ref{3.2}). At the left-hand
side we may invoke the perfect-screening rule (\ref{3.7}) for $\ell=1,2$,
when the sum over $\sigma_1$ with weights $n_{\sigma_1}\; e_{\sigma_1}$
is carried out as well. In this way we arrive at the identity:
\begin{eqnarray}
\fl (d-2)\; \beta \sum_{\sigma_1,\sigma_2,\sigma_3} n_{\sigma_1}\; n_{\sigma_2}\;
n_{\sigma_3}\; e_{\sigma_1}\; e_{\sigma_2}\; e_{\sigma_3}
\int d{\bf r}_2\; d{\bf r}_3 \; 
h^{(3)}_{\sigma_1\sigma_2\sigma_3}({\bf r}_1,{\bf r}_2,{\bf r}_3)\;
r^2_{23}\; \phi(r_{12})\nonumber\\
+ (d-2)\; \beta \sum_{\sigma_1,\sigma_2} n_{\sigma_1}\; n_{\sigma_2}\;
 e^2_{\sigma_1}\; e_{\sigma_2}
\int d{\bf r}_2\; h^{(2)}_{\sigma_1\sigma_2}({\bf r}_1,{\bf r}_2)\;
r^2_{12}\; \phi(r_{12})=\nonumber\\
=\frac{d}{2(d+2)}\; \beta\; q_v
\sum_{\sigma_1,\sigma_2} n_{\sigma_1}\; n_{\sigma_2}\;
 e_{\sigma_1}\; e_{\sigma_2}
\int d{\bf r}_2\; h^{(2)}_{\sigma_1\sigma_2}({\bf r}_1,{\bf r}_2)\; r^4_{12}\nonumber\\
-2\; d\; \sum_{\sigma_1,\sigma_2} n_{\sigma_1}\; n_{\sigma_2}\;
 e_{\sigma_1}
\int d{\bf r}_2\; h^{(2)}_{\sigma_1\sigma_2}({\bf r}_1,{\bf r}_2)\;
r^2_{12}\nonumber\\
-(d-2)\; c_d\; \beta\sum_{\sigma_1,\sigma_2} n_{\sigma_1}\; n_{\sigma_2}\;
e^2_{\sigma_1}\; e_{\sigma_2} \int d{\bf r}_2\;  
h^{(2)}_{\sigma_1\sigma_2}({\bf r}_1,{\bf r}_2)\; r^2_{12}
\label{C.4}
\end{eqnarray}
which is needed in section 6 of the main text.

Finally, we want to establish an equality for the four-particle Ursell
function. It follows by starting from an equality for $h^{(4)}$ of a
similar form as (\ref{C.1}):
\begin{eqnarray}
\fl (d-2)\sum_{\sigma_3,\sigma_4}n_{\sigma_3}\;n_{\sigma_4}\;e_{\sigma_3}\;e_{\sigma_4}
\int d{\bf r}_3\; d{\bf r}_4\; 
h^{(4)}_{\sigma_1\sigma_2\sigma_3\sigma_4}({\bf r}_1,{\bf r}_2,{\bf
  r}_3,{\bf r}_4)\; 
\left[\phi(r_{34})-c_d\right]  =\nonumber\\
\fl =2\sum_{\sigma_3,\sigma_4}n_{\sigma_3}\;n_{\sigma_4}\;e_{\sigma_3}\;e_{\sigma_4}
\int d{\bf r}_3\; d{\bf r}_4\; 
h^{(4)}_{\sigma_1\sigma_2\sigma_3\sigma_4}({\bf r}_1,{\bf r}_2,{\bf
  r}_3,{\bf r}_4)\; 
{\bf r}_{13}\cdot\frac{\partial \phi(r_{34})}{\partial {\bf r}_3}
\label{C.5}
\end{eqnarray}
Upon using the hierarchy equation (\ref{3.1}) for $k=3$, the expansion
(\ref{3.3}), the identity (\ref{3.4}) and the perfect-screening rules
(\ref{3.5})--(\ref{3.6}) we get, by taking steps analogous to those of
appendix B of ref.\ \cite{SvW87}:
\begin{eqnarray}
\fl \half\;(d-2)\; \beta\sum_{\sigma_3,\sigma_4} 
n_{\sigma_3}\;n_{\sigma_4}\;e_{\sigma_3}\;e_{\sigma_4}
\int d{\bf r}_3\; d{\bf r}_4\; 
h^{(4)}_{\sigma_1\sigma_2\sigma_3\sigma_4}
({\bf r}_1,{\bf r}_2,{\bf r}_3,{\bf r}_4)\; \phi(r_{34})=\nonumber\\
\fl =-(d-2)\; \beta\sum_{\sigma_3} n_{\sigma_3}\;e_{\sigma_3}\int d{\bf r}_3\;
h^{(3)}_{\sigma_1\sigma_2\sigma_3}({\bf r}_1,{\bf r}_2,{\bf r}_3)
\; \left[e_{\sigma_1}\; \phi(r_{13})+e_{\sigma_2}\;
  \phi(r_{23})\right]\nonumber\\
\fl +\half\;\beta\; q_v\sum_{\sigma_3}n_{\sigma_3}\;e_{\sigma_3}
\int d{\bf r}_3\;
h^{(3)}_{\sigma_1\sigma_2\sigma_3}({\bf r}_1,{\bf r}_2,{\bf r}_3)\;
r^2_{23}\nonumber\\
\fl -d\sum_{\sigma_3}n_{\sigma_3}\int d{\bf r}_3\;
h^{(3)}_{\sigma_1\sigma_2\sigma_3}({\bf r}_1,{\bf r}_2,{\bf r}_3) 
 +{\bf r}_{12}\cdot\frac{\partial}{\partial{\bf r}_1}\; 
h^{(2)}_{\sigma_1\sigma_2}({\bf r}_1,{\bf r}_2)\nonumber\\
\fl -\beta \sum_{\sigma_3,\sigma_4}
n_{\sigma_3}\;n_{\sigma_4}\;e_{\sigma_3}\;e_{\sigma_4}
\int d{\bf r}_3\; h^{(2)}_{\sigma_1\sigma_3}({\bf r}_1,{\bf r}_3)\;
{\bf r}_{13}\cdot\frac{\partial \phi(r_{23})}{\partial{\bf r}_2}
\int_{r_{24}>r_{23}} d{\bf r}_4\; 
h^{(2)}_{\sigma_2\sigma_4}({\bf r}_2,{\bf r}_4)\nonumber\\
\fl +(d-2)\; \beta\sum_{\sigma_3,\sigma_4}
n_{\sigma_3}\;n_{\sigma_4}\;e_{\sigma_3}\;e_{\sigma_4}
\int d{\bf r}_3\; h^{(2)}_{\sigma_2\sigma_3}({\bf r}_2,{\bf r}_3)\;
\left[\phi(r_{13})-c_d\right]\nonumber\\
\times \int_{r_{14}>r_{13}} d{\bf r}_4\; 
h^{(2)}_{\sigma_1\sigma_4}({\bf r}_1,{\bf r}_4)\nonumber\\
\fl +(d-2)\; \beta\; e_{\sigma_2}\; \left[\phi(r_{12})-c_d\right]\sum_{\sigma_3} 
 n_{\sigma_3}\;e_{\sigma_3}\int_{r_{13}>r_{12}} d{\bf r}_3\;
h^{(2)}_{\sigma_1\sigma_3}({\bf r}_1,{\bf r}_3)\nonumber\\
\fl -(d-2)\; \beta\; e_{\sigma_1}\; e_{\sigma_2}\;
h^{(2)}_{\sigma_1\sigma_2}({\bf r}_1,{\bf r}_2)\;
\phi(r_{12})
+\half\;\beta\; q_v\; e_{\sigma_1}\; 
h^{(2)}_{\sigma_1\sigma_2}({\bf r}_1,{\bf r}_2)\; r^2_{12}\nonumber\\
\fl -\half\; (d-2)\; c_d\; \beta\; \left[
\sum_{\sigma_3} n_{\sigma_3}\; e^2_{\sigma_3}\int d{\bf r}_3\;
h^{(3)}_{\sigma_1\sigma_2\sigma_3}({\bf r}_1,{\bf r}_2,{\bf r}_3) 
+\left( e^2_{\sigma_1}+e^2_{\sigma_2}\right)\; 
h^{(2)}_{\sigma_1\sigma_2}({\bf r}_1,{\bf r}_2)\right]\nonumber\\
\label{C.6}
\end{eqnarray}
This rather complicated identity has been used in appendix B. Inspection of the
terms at the right-hand side shows that for large $r_{12}$ each of these
vanishes faster than any inverse power of $r_{12}$, as it should be in view
of the short-range character of the four-point Ursell function at the
left-hand side.

For $d>2$ the above identities may be simplified by putting $c_d$ equal to
0. That choice is not allowed when one is interested in the limit
$d\rightarrow 2$. In that case one takes $c_d$ according to
(\ref{2.3}). In the limit $d\rightarrow 2$ the left-hand side of the
identity (\ref{C.2}) vanishes, so that we get an identity that
connects the zeroth and second moments of the two-particle Ursell function:
\begin{eqnarray}
\fl 
\beta\; q_v\sum_{\sigma_2}n_{\sigma_2}\; e_{\sigma_2}\int d{\bf r}_2\; 
h^{(2)}_{\sigma_1\sigma_2}({\bf r}_1,{\bf r}_2) \; r_{12}^2
=4\sum_{\sigma_2}n_{\sigma_2}\int d{\bf r}_2\; 
h^{(2)}_{\sigma_1\sigma_2}({\bf r}_1,{\bf r}_2)\nonumber\\
-\frac{\beta}{2\pi}\left[\sum_{\sigma_2}n_{\sigma_2}\;
  e_{\sigma_2}^2 \int d{\bf r}_2\; h^{(2)}_{\sigma_1\sigma_2}({\bf r}_1,{\bf r}_2)
+ e_{\sigma_1}^2\right]
\label{C.7}
\end{eqnarray}
It should be noted that the last two terms would have been missed when in
(\ref{C.2}) the limit $d\rightarrow 2$ had been taken naively after putting $c_d=0$. 
The identity (\ref{C.7}), which is valid for the special case $d=2$ only,
has been obtained recently \cite{JS08}. Upon substituting (\ref{5.3}) in
the right-hand side and using (\ref{A.10}) we recover (\ref{5.8}) for
$d=2$. In fact, this shows that for $d=2$ the second-moment sum rule
(\ref{5.8}) can be derived from perfect screening, symmetry and
thermodynamics alone, without having recourse to the rather complicated
expression for the derivative $\partial n_\sigma/\partial\beta$ of the
partial density with respect to the inverse temperature. The latter
expression is essential in deriving the second-moment sum rule for
arbitrary $d>2$.

Similarly, for $d \rightarrow 2$ the symmetry relation (\ref{C.4}) reduces
to an identity connecting the second and fourth moments of the two-particle
Ursell function:
\begin{eqnarray}
\fl \beta\; q_v
\sum_{\sigma_1,\sigma_2} n_{\sigma_1}\; n_{\sigma_2}\;
 e_{\sigma_1}\; e_{\sigma_2}
\int d{\bf r}_2\; h^{(2)}_{\sigma_1\sigma_2}({\bf r}_1,{\bf r}_2)\; r^4_{12}=\nonumber\\
=16 \sum_{\sigma_1,\sigma_2} n_{\sigma_1}\; n_{\sigma_2}\;
 e_{\sigma_1}
\int d{\bf r}_2\; h^{(2)}_{\sigma_1\sigma_2}({\bf r}_1,{\bf r}_2)\;
r^2_{12}\nonumber\\
-\frac{2\beta}{\pi} \sum_{\sigma_1,\sigma_2} 
n_{\sigma_1}\; n_{\sigma_2}\; e^2_{\sigma_1} \; e_{\sigma_2}
\int d{\bf r}_2\; h^{(2)}_{\sigma_1\sigma_2}({\bf r}_1,{\bf r}_2)\;
r^2_{12}
\label{C.8}
\end{eqnarray}
The last term is missed when one puts $c_d=0$ in (\ref{C.4}) before taking
the limit $d\rightarrow 2$. Substituting (\ref{5.8}) and using the equation
of state (\ref{4.7}) we are led to (\ref{6.6}) for $d=2$. Hence, in a
similar way as discussed above for the second-moment sum rule, the
derivation of the fourth-moment sum rule can be simplified for the special
case $d=2$. For that case it is enough to make use of the perfect-screening
and second-moment rules, symmetry properties and thermodynamical relations
in the proof, whereas for general $d$ the derivative $\partial
h^{(2)}_{\sigma_1\sigma_2}/\partial \beta$ of the two-particle Ursell
function with respect to the inverse temperature needs to be
determined. Incidentally, we remark that for the one-component case the
identity (\ref{C.8}) has been obtained before \cite{V87}.

We are left with (\ref{C.6}) in the limit $d\rightarrow 2$. The resulting identity
is rather complicated and is not needed in the main text, so that we
refrain from writing it down.

\end{document}